\documentstyle[12pt,frascatiphys,epsfig]{article}
\def\ompipi  {  \omega \pi^- \pi^0  }
\def\omrho   {  \omega \rho^-  }
\def\reactf  { $\pi^- Be \rightarrow \pi^+ 2 \pi^- 2 \pi^0 Be$ }

\def\rom2pi  { \pi^- Be \rightarrow \omega \pi^- \pi^0 Be }
\def\at      {  a_2 (1320)  }
\def\aq      {  a_4 (2040)  }
\def\gev     { GeV }
\def\mev     { MeV }
\def\tprime  { t^{\prime} }
\def\tripi   { \pi^+\pi^-\pi^0 }
\def\tripc   { \pi^+\pi^-\pi^- }

\def\jpme    { J^PM^{\eta}}
\begin{document}
\title{ 
NEW RESULTS FROM VES. 
}
\author{
 Valeri Dorofeev, VES collaboration \\{\em IHEP, Protvino, Russia, RU-142284}
}
\maketitle
\baselineskip=14.5pt
\begin{abstract}
  The results of the partial wave analysis(PWA) of the 
$\pi^+\pi^-\pi^-$ and $\ompipi$ systems are presented. 
The $a_3$ and $a_4(2040)$
signals are observed in the  $\rho(770)\pi$ and $f_2(1270)\pi$ channels.
Indications of the $a_1'$ meson existence was found in the $1^+0^+$ $\rho\pi$ S-wave.
The decay branching ratio of the $ \at^- $ to $\ompipi$ was measured.
The $2^+1^+$ wave shows a broad bump at $ M \approx 1.7 \, \gev $. 
The decays of the $\pi_2(1670)$, $ \aq$ and $\pi(1740)$ into $ \omrho $ were found. 
The  resonance in the $b_1(1235)\pi$ wave with exotic quantum numbers $J^{PC}=1^{-+}$
at $ M \approx 1.6 \, \gev $ is observed and the simultaneous analysis
of the $1^{-+}$ wave in the  $b_1(1235)\pi$, $\eta '\pi$ and $\rho\pi$
final states is presented.  
\end{abstract}
\baselineskip=17pt
\begin{figure}[htb]
\epsfig{file=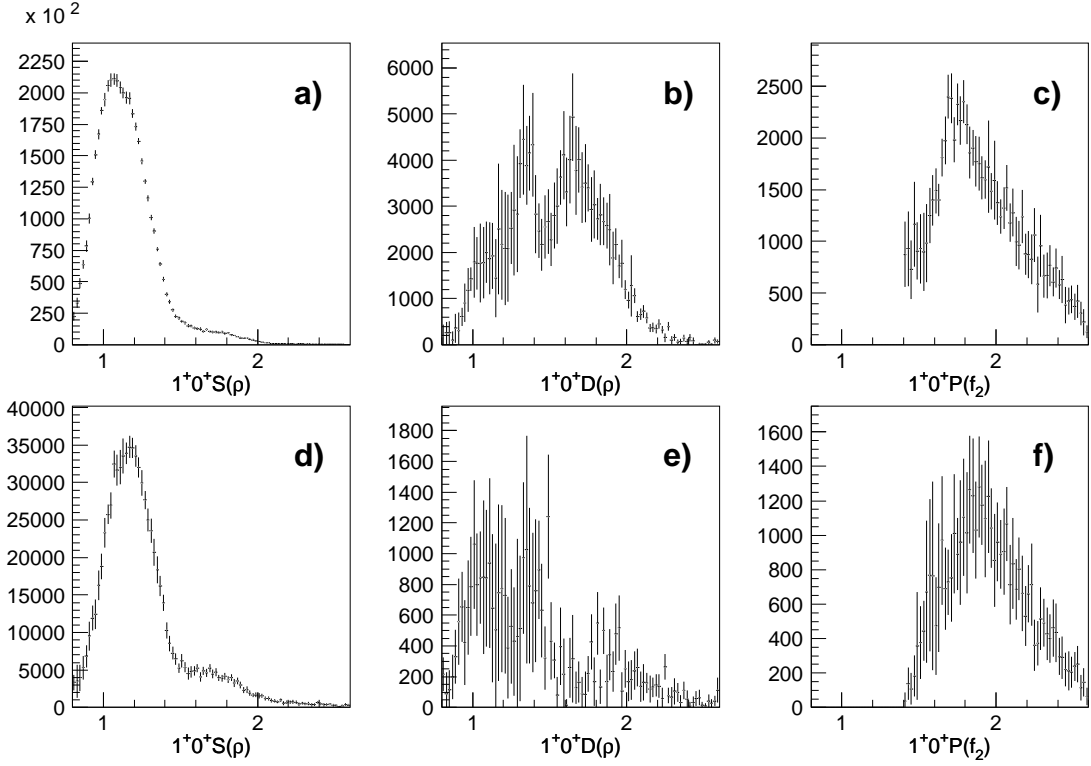,width=\textwidth,%
bbllx=0pt,bblly=0pt,bburx=420pt,bbury=280pt}
\caption{Main $J^{PC} = 1^{++}$ waves
for $t'<0.06 \, \gev^2$ (a-c) and $0.06<t'<0.7 \, \gev^2$ (d-f)}
\label{fig2}
\end{figure}

\begin{figure}[htb]
\epsfig{file=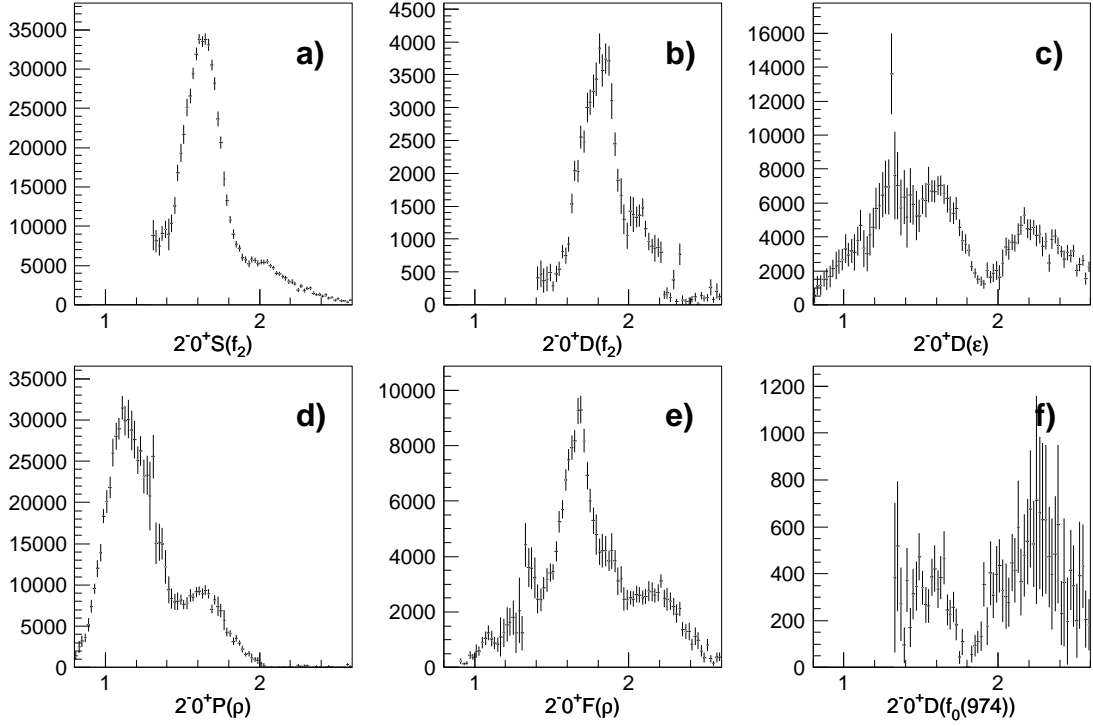,width=\textwidth,%
bbllx=0pt,bblly=0pt,bburx=420pt,bbury=280pt}
\caption{Main $J^{PC} = 2^{-+}$ waves for $t'<0.06 \, \gev^2$}
\label{fig3}
\end{figure}

\begin{figure}[htb]
\epsfig{file=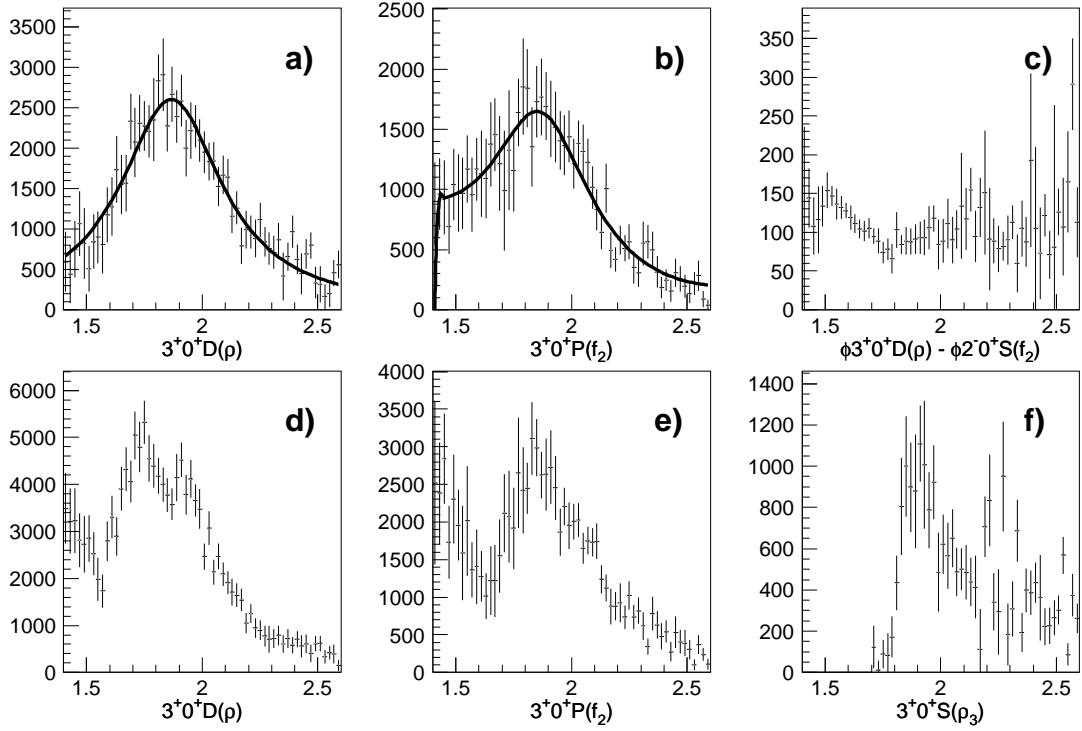,width=\textwidth,%
bbllx=0pt,bblly=0pt,bburx=420pt,bbury=280pt}
\caption{The $3^+0^+D\rho$ (a), $3^+0^+P f_2$ (b) wave
intensities, the phase difference
$\phi(3^+D\rho) - \phi(2^-S f_2)$ (c) for $0.06<\tprime<0.7 \, \gev^2$;
the  $3^+0^+D\rho$ (d), $3^+0^+P f_2$(e) and $3^+0^+S\rho_3$(f)
 wave intensities for $\tprime < 0.06 \, \gev^2$ }
\label{fig4}
\end{figure}

\begin{figure}[htb]
\epsfig{file=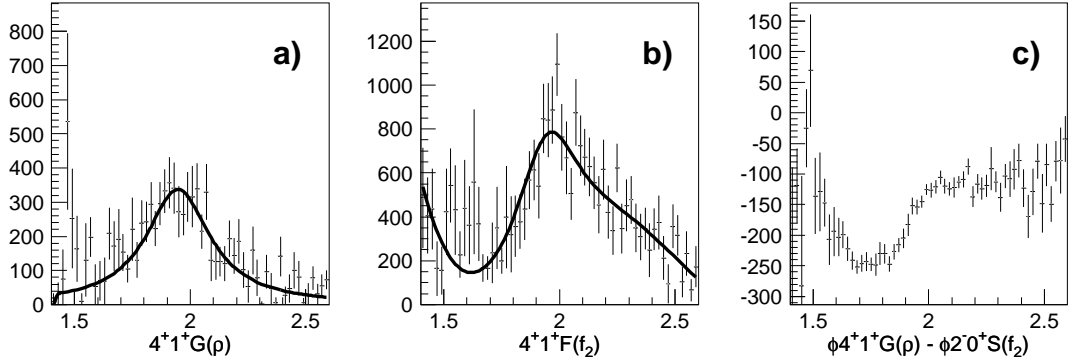,width=\textwidth,%
bbllx=0pt,bblly=125pt,bburx=420pt,bbury=280pt}
\caption{The $4^+1^+D\rho$ (a) and  $4^+1^+F f_2$ (b) wave
intensities and the phase difference
$\phi(4^+G\rho) - \phi(2^-S f_2)$ for $0.06 < \tprime < 0.7 \, \gev^2$.}
\label{fig5}
\end{figure}

\begin{figure}[htb]
\epsfig{file=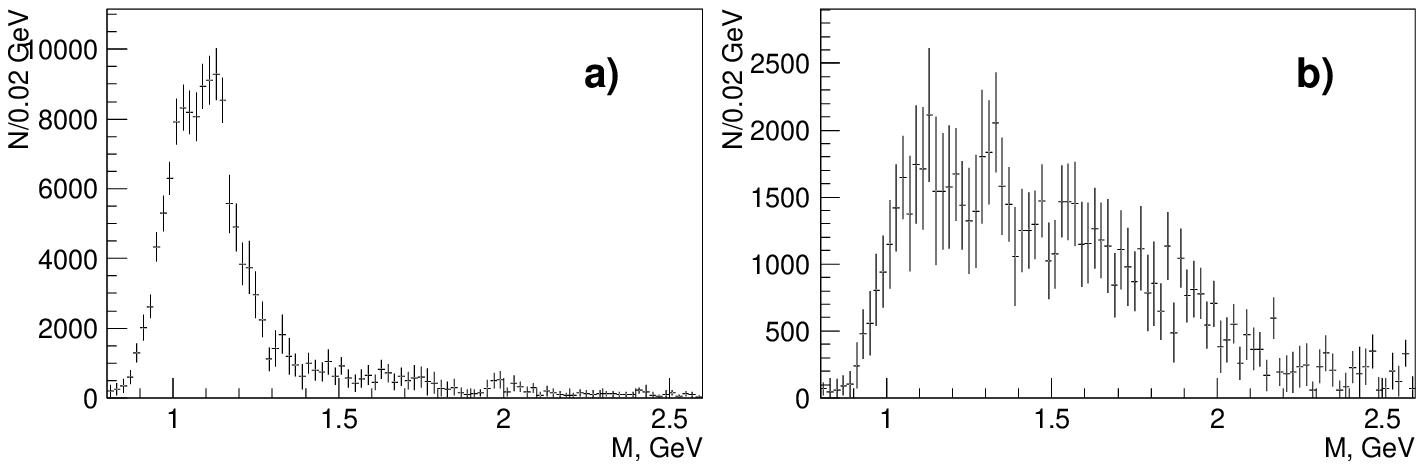,width=\textwidth,%
bbllx=0pt,bblly=130pt,bburx=420pt,bbury=280pt}
\caption{The $\jpme=1^-1^+S\rho$ wave intensity for 
$\tprime< 0.06 \, \gev^2$ (a) and for $0.06 < \tprime < 0.7 \, \gev^2$ (b).}
\label{fig6}
\end{figure}

\section{Introduction}
We present the results of the PWA of $ \tripc$, produced in 
reaction: 
\begin{eqnarray}
\label{ppp}     \pi^- Be & \to & \pi^+ \pi^- \pi^- Be
\end{eqnarray}
and $\ompipi$ system, produced in: 
\begin{equation}
\rom2pi , \quad \omega \rightarrow \tripi
\label{rom2pi}  
\end{equation}
Our previous results of the analysis of reaction (\ref{ppp})
were published in \cite{bec3,maryland,pippm}, and of reaction (\ref{rom2pi})
were partially reported at the conferences \cite{Manchester},
 \cite{Warsaw}.
The measurements were carried out using VES spectrometer exposed
by the $37 \, \gev$ momentum $\pi^-$ beam.
The description of the setup can be found in \cite{bec1}.

\section{ Results of the $\tripc$-system PWA. }
The selection criteria for reaction (\ref{ppp}) 
and the description of the PWA procedure can be found in \cite{pippm}.
The relativistic covariant
helicity formalism \cite{ChungAmp} is used 
to construct the amplitudes and the positively definite density matrix
of the full rank.
The largest waves of the $J^P=0^-$, $1^+$, $2^-$ channels
are decoupled from the other waves with the same $J^PM^{\eta}$
and are free to interfere with each other. 
The $6\cdot10^6$ events 
with $|t'| < 0.06 \, \gev^2$ 
\footnote{ $\tprime = t - t_{min}$, where $t$- momentum transfer from the beam to the
final state squared, $t_{min}$-its minimum value.}
and $2\cdot10^6$ with 
$0.06 < |t'| < 0.7 \, \gev^2$ are analyzed separately.

We present the main features of the most significant waves in the
high $3\pi$ mass region.

$\jpme = 1^+0^+$.
A peak in the $1^+0^+D\rho$ wave (Fig.\ref{fig2}(b)) 
and the shoulder in the $1^+0^+S\rho$ wave (Fig.\ref{fig2}(a), (d)) are observed
at $M\approx 1.7 \, \gev$ and are considered to be
the $a_1'(1700)$ decay into the $\rho\pi$.
The peak was fitted with the coherent sum of the Breit-Wigner resonance and the
exponential background.   
The fit results in the following $a_1'(1700)$  parameters:
$$
\begin{array}{ll} 
M=1.80 \pm 0.05 \, \gev & \Gamma=0.23^{+0.10}_{-0.03} \, \gev ; \\
\end{array}
$$
where the errors are dominated by systematics. 
The $a_1'(1700)$ branching ratios into the D-wave $\rho\pi$ and P-wave
$f_2(1270)\pi$ are found to be:
$$
\frac{Br(a_1'(1700) \to (\rho\pi)_D)}
     {Br(a_1'(1700) \to (\rho\pi)_S)} < 0.35 \quad
\frac{Br(a_1'(1700) \to f_2\pi)}
     {Br(a_1'(1700) \to (\rho\pi)_S)} < 0.23 \mbox{ at 95\% CL}
$$

  $\jpme = 2^-0^+$. A complicated wave behaviour for low $\tprime$
shown in Fig.\ref{fig3} is described by the interplay
of the $\pi_2(1670)$ and $\pi_2'(2100)$ states \cite{pippm,daum}.
A peak at $M \approx 1.7 \, \gev$ in the F-wave $\rho\pi$ wave intensity
is observed.

   $\jpme = 3^+0^+$.  A resonance-like signal is observed near 
$M\approx 1.8\, \gev$ in 
the $\rho\pi$ and $f_2\pi$ waves, see Fig.\ref{fig4}.
The phase of the $3^+0^+D\rho$ as related to the $2^-0^+Sf_2$ 
is shown in Fig.\ref{fig4}(c) and is in accordance with the expectations
for a resonance. The simultaneous fit of the $f_2\pi$ 
intensity with the relativistic Breit-Wigner function and the $\rho\pi$ 
intensity with the incoherent sum of the Breit-Wigner and the Chebyshev polynomial 
background results in the following parameters of the resonance:
$$
\begin{array}{ll} 
    M=1.86  \pm  0.02\, \gev, & \Gamma  =  0.54 \pm  0.03\, \gev. \\
\end{array}
$$
The relative probability of decay into the $f_2(1270)\pi$ and $\rho(770)\pi$ is
as follows:
\begin{displaymath}
 \frac{Br(a_3 \to f_2(1270)\pi)}
          {Br(a_3 \to \rho\pi)}
   = 0.5 \pm 0.1. \nonumber
\end{displaymath}

   $\jpme=4^+1^+$.  
A bump near $M \approx 2 \, \gev$ is found in the $4^+1^+G\rho$ and
$4^+1^+F f_2$ waves produced at high $\tprime$ (Fig.\ref{fig4}).
The phase of the $4^+1^+G\rho$ relative to the $\pi_2(1670)$ 
is in accordance with the expectation for the resonance.
The simultaneous fit of the $f_2\pi$ 
intensity with the relativistic Breit-Wigner function and the $\rho\pi$ 
intensity with the incoherent sum of the Breit-Wigner and the Chebyshev 
polynomial background results in the following parameters of the resonance:
$$
\begin{array}{ll} 
    M       =  1.95  \pm 0.02\, \gev, &  \Gamma  =  0.34   \pm 0.10\, \gev. \\
\end{array}
$$
We identify this object with the $\aq$, having the following branching ratio:
\begin{eqnarray}
\frac{Br( \aq \rightarrow f_2(1270)\pi )}{Br( \aq \rightarrow \rho\pi)}=
0.5 \pm 0.2. \nonumber 
\end{eqnarray}

 $\jpme = 1^-1^+$. 
The observation of a 
signal in the $1^-1^+P(\rho)$ wave with
$M = 1.62 \pm 0.02 \, \gev$, $\Gamma = 0.24 \pm 0.05 \, \gev$ 
was previously reported as preliminary by the VES \cite{GouzD}.
Later, the observation of the state with
$M = 1593 \pm 8 ^{+20}_{-47} \, \mev$, $\Gamma = 168 \pm 20 ^{+150}_{-12} \, \mev$  
was declared by the E852 \cite{ostro}. It was shown in \cite{zaitsev} the
model dependence of the signal behaviour.
We do not observe such a narrow signal (Fig.~\ref{fig6}) in the fit results
with the applied PWA model.
However there appears a peak at $M \approx 1.6 \, \gev$ with 
$\Gamma \approx 0.3 \, \gev$ in the $1^-1^+$ wave intensity, described
by the leading term in the expansion of the density matrix in terms of the 
eigenvalues ( shown in Fig.\ref{fig0}).
%
\begin{figure}[htb]
\epsfig{file=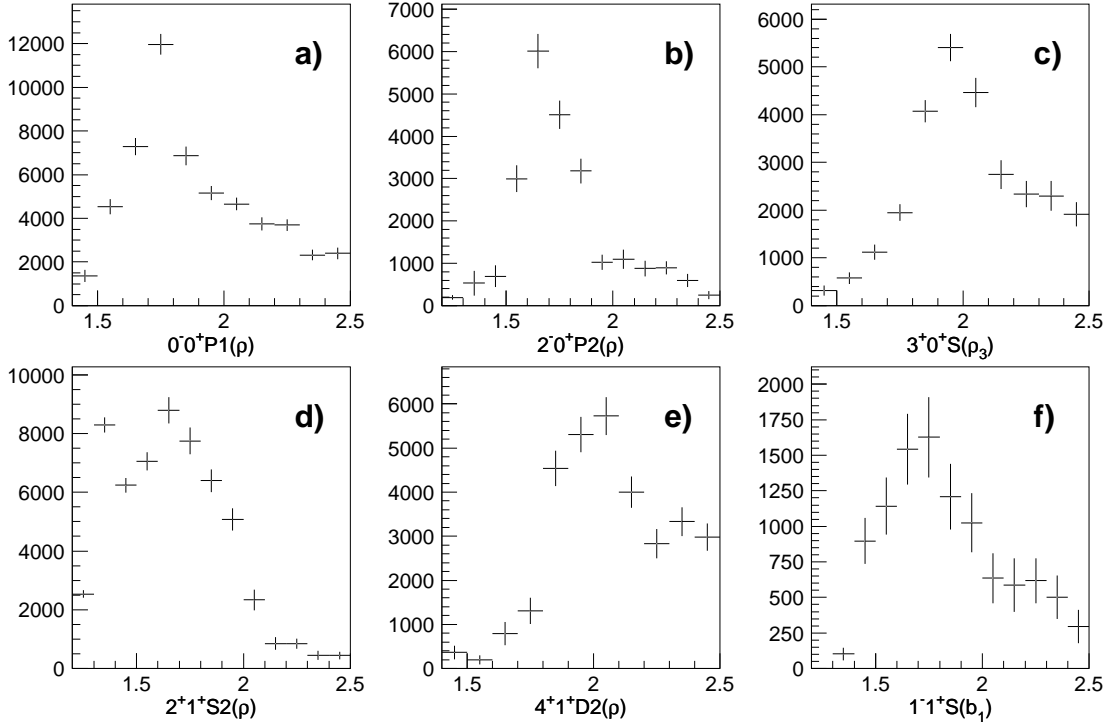,width=\textwidth,%
bbllx=0pt,bblly=0pt,bburx=560pt,bbury=380pt}
\caption{Wave intensities of:
a) $0^-0^+P1(\rho)$, 
b) $2^-0^+P2(\rho)$ with $-\tprime < 0.08 \, \gev^2 $.,
Wave intensities of: 
c) $3^+0^+S(\rho_3)$, 
d) $2^+1^+S2(\rho)$, 
e) $4^+1^+D2(\rho)$, 
f) $1^-1^+S(b_1)$.
}
\label{waves_tot}
\end{figure}
\section{The results of the $\ompipi$-system PWA} 
The selection criteria for reaction (\ref{rom2pi}) 
and the description of the PWA procedure can be found in \cite{Dorofeev}.
The results of the PWA are presented in Fig.~\ref{waves_tot}.

$\jpme=0^-0^+$.
  A peak in the region of $1.74 \, \gev$ with the flat background 
dominates in the wave  $0^-0^+P(\rho)$ at low $\tprime$ 
(Fig.~\ref{waves_tot}(a)). 
The phase of this wave relative to the smooth $1^+0^+P(b_1)$ wave has the
resonant behaviour. 
The resonance parameters were determined by the fit of the wave intensity with
the incoherent sum of a relativistic Breit-Wigner function and a cubic 
polynomial and are as follows:
the mass $M=1.737 \pm 0.005 \pm 0.015 \, \gev$ and the width  
$\Gamma=0.259 \pm 0.019 \pm 0.06 \, \gev$.

$\jpme=2^-0^+$.
A clear peak is observed
at  $ M \sim 1.67 \, \gev$ with $ \Gamma \sim 0.2 \, \gev$
(Fig.~\ref{waves_tot}(b)). 
The resonant phase behaviour of the $2^-0^+P1(\rho)$ and $ 2^-0^+P2(\rho)$
waves relative to the $1^+0^+P(b_1)$ wave is observed.
The resonance parameters of the
$2^-0^+P2(\rho)$ peak were estimated in the same way as for $\pi(1740)$:
 $M=1.687 \pm 0.009 \pm 0.015 \, \gev \mbox{ and } 
\Gamma=0.168 \pm 0.043 \pm 0.053 \, \gev$. We identify this 
phenomenon with the decay of the $\pi_2(1670)$ into $\omega\rho$.
The partial branching ratio was found by normalization to
the decay $\pi_2(1670) \rightarrow f_2(1270)\pi$ \cite{pippm},
observed in the current experiment decay: 
\begin{eqnarray}
Br(\pi_2(1670)^- \rightarrow \omrho ) = 0.027\pm0.004\pm 0.01 \nonumber 
\end{eqnarray}
The limits of the $\pi_2(1670)$ decay branching ratios are set 
at the $2\sigma$ confidence level:
$$
\begin{array}{ll}
Br(\pi_2(1670) \rightarrow \rho_1(1450)\pi)<0.0036 , &
Br(\pi_2(1670) \rightarrow b_1 \pi)<0.0019 . \\
\end{array}
$$

$\jpme=3^+0^+$.
A peak at $M_{5\pi} \sim 2 \, \gev$  and $ \Gamma \sim 0.35 \, \gev $ is clearly
seen in the $3^+0^+S(\rho_3(1690))$ wave for events with low $\tprime$
(Fig.~\ref{waves_tot}(c)). 
However the resonant phase motion was not found. Such wave behaviour can be
attributed to the Deck effect process \cite{Deck}. 

$\jpme=2^+1^+$.
The intensive production and
decay of the $\at$ is the main process at low masses (Fig.~\ref{waves_tot}(d)). 
The decay probability was found to be $Br(\at \rightarrow \ompipi)=( 5 \pm 1 ) \% $.
We define the $\at$ partial width
as that of a Breit-Wigner function with the S-wave $\ompipi$ background. 
The nature of the $1.7 \, \gev$ mass structure is unknown. 
There may be another resonance or the opening of the $\omega\rho$ channel \cite{Migdal}.
We can not make preference to a particular hypothesis.

$\jpme=4^+1^+$.
The signal at $M \approx 2 \, \gev$ (Fig.~\ref{waves_tot}(e)) with the
resonant phase behaviour can be identified as the $\aq$. 
The $\aq$ parameters are estimated by
the fit with the incoherent sum of a $D$-wave relativistic Breit-Wigner function
and a polynomial background to be:     
 $M=1.944 \pm 0.008 \pm 0.050 \, \gev \mbox{ and } \Gamma=0.324 \pm 0.026 \pm 0.075 \, \gev$.
The $\tprime$-dependence of the $\aq$ is identical to that of the $\at$.

$\jpme=1^-1^+$.
The intensity of the $1^-1^+S(b_1)$ wave for events 
with high $\tprime$ shows a wide bump with maximum at  
$M \sim 1.6 \div 1.7 \, \gev$. 
The highest intensity of this wave does not exceed $15 \%$ of the  
$2^+1^+S2(\rho)$ wave intensity (Fig.~\ref{waves_tot}(f)).
The $\omega\rho$ P-waves are included in turn along with the $b_1\pi$ wave.
Their intensity distribution differs in form from that of
the $b_1\pi$ wave and their inclusion in the fit do not influence on the
$b_1\pi$ wave intensity behaviour.
\begin{figure}[htb]
\epsfig{file=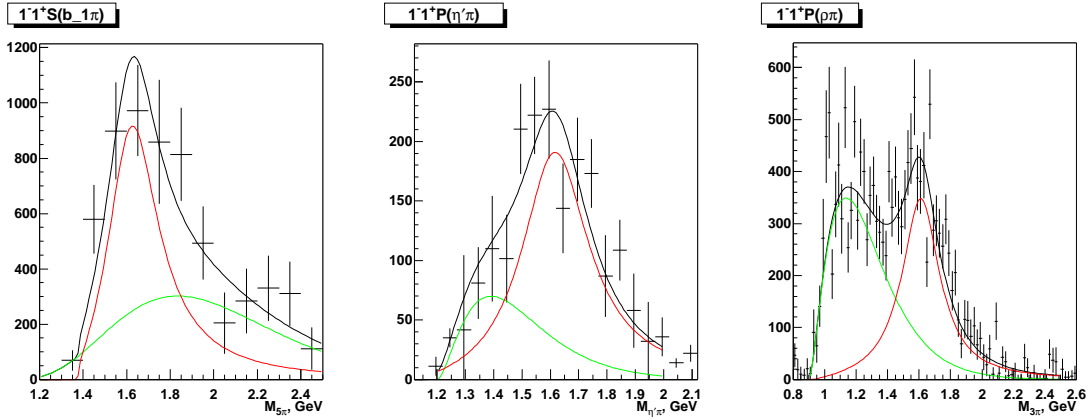,width=\textwidth,%
bbllx=0pt,bblly=200pt,bburx=568pt,bbury=440pt}
\caption{Intensities of the $\jpme = 1^-1^+$ waves in the channels:
$b_1(1235)\pi$, $\eta'\pi$ and $\rho\pi$.}
\label{fig0}
\end{figure}
\section{The $\jpme=1^-1^+$ wave analysis} 
The combined analysis of the $\jpme=1^-1^+$
$b_1\pi$ and the $\jpme=2^+1^+$ $\omega\rho$ was carried out in order to
understand the nature of the $b_1\pi$ wave. The results of the $\ompipi$
system PWA were used for this analysis. The diagonal elements  
and the real and imaginary parts of the non-diagonal element of the 
$\rho$-matrix corresponding to the $1^-1^+$ $b_1\pi$ and $2^+1^+$ 
$\omega\rho$ waves with high $\tprime$ were simultaneously fitted.
Fit results were used to predict the coherence parameter for cross checking.
The $b_1\pi$ amplitude was saturated by the Breit-Wigner resonance
and the coherent background. The $\omrho$ amplitude was saturated 
by the $a_2(1320)$-meson, a background and an $a_2'$ state has been also tried. 
The results of the fits with various ways of the $\omrho$ wave construction
point out to the resonance nature of the $b_1\pi$ signal. 
The range of the parameters variation is large due to the freedom in the 
$2^+1^+$ state model.   
\par
 The signal in the 
$\jpme=1^-1^+$ wave of the $\eta'\pi^-$-system with close parameters was
observed  earlier \cite{Gouz2}.
The simultaneous fit of the $b_1\pi$ and $\eta'\pi$ intensities 
with incoherent sum of a Breit-Wigner resonance and a
background in each channel was carried out
( Fig.~\ref{fig0}). The fit results in the following
parameters:
$$
\begin{array}{ll} 
    M       =  1.58  \pm 0.03 \, \gev, &  \Gamma  =  0.30  \pm 0.03 \, \gev. \\
\end{array}
$$
The form of the signal in the $1^-1^+$ $\rho\pi$ wave is close to the Breit-Wigner
function with these parameters. A fit of all three channels results in the resonance
parameters changed within errors.
All these facts indicate to the existence of a wide resonance 
$\Gamma = 0.29 \pm 0.03  \, \gev$ with the mass $M = 1.61 \pm 0.02  \, \gev $ and
the relative branching ratio:
$
Br(b_1\pi) : Br(\eta'\pi) : Br(\rho\pi) = 1 : 1.0\pm0.3 : 1.6\pm0.4. 
$
\section{Conclusions} 
The PWA of the reaction $\pi^-Be \to \pi^+\pi^-\pi^-Be$ 
and \reactf was performed.
\par
The mass and the width 
of the resonance structure in the $\jpme=0^- 0^+ $  $\omrho$ wave
differs from that for $\pi(1800)$.
There may exist two objects of the different nature: a hybrid $\pi(1800)$ and a 
$3^1S_0 \mbox{ } q \bar q$ state decaying into $\omega\rho$.
\par
The indication of the existence of the
resonance $a_1'$ mostly decaying to the $\rho\pi$ in the S-wave is found. 
\par
The $\pi_2(1670)$ decays into the $ \omega \rho$ and
$\rho\pi$ in the F-wave are found.
\par
The $\at^- \rightarrow \ompipi $ decay and
a wide bump of unknown nature at $M \approx 1.7 \, \gev$ are observed
in the $ \jpme=2^+1^+$ wave.
The $a_3$ and $\aq$ decays to the $\rho\pi$ and $f_2\pi$ are observed with
the following relative branching ratio of the $ a_4(2040)$ decays:
$
 Br(f_2(1270)\pi) : Br(\rho\pi) : Br(\omega\rho) = 0.5\pm0.2 : 1 : 1.5\pm0.4 
$
\par
The preliminary results of the $1^{-+}$ wave analysis point out to the
existence of the resonance with
exotic quantum numbers, which are forbidden for $q \bar q$ states,
the mass $M=1.61 \pm 0.02 \, \gev$, the width
$\Gamma=0.29 \pm 0.02 \, \gev$ and the following relative branching ratio:
$
Br(b_1\pi) : Br(\eta'\pi) : Br(\rho\pi) = 1 : 1.0\pm0.3 : 1.6\pm0.4. 
$
%
%

\end{document}